\documentclass{PoS}

\usepackage{ragged2e}
\usepackage{color}
\usepackage{url}
\usepackage{relsize}
\usepackage{float}

\title{Detection of vertical muons with the HAWC water Cherenkov detectors and it's application to gamma/hadron discrimination}

\ShortTitle{Vertical muons detection in HAWC regarding a new G/H separation strategy}

\author{Ariel Z\'{u}\~{n}iga-Reyes$^{a,}$\footnote{Corresponding author: \color{blue}arzure89@gmail.com} , Alberto Hern\'{a}ndez$^{a, b}$, Aldo Miranda-Aguilar$^{c}$, Andr\'{e}s Sandoval$^{a}$, J. Mart\'{i}nez-Castro$^{c}$, Rub\'{e}n Alfaro$^{a,}$\footnote{Presenter} , Ernesto Belmont$^{a}$, Hermes Le\'{o}n$^{a}$, Ana P. Vizcaya$^{a}$ for the HAWC collaboration$^{d}$\\

$^{a}$Instituto de F\'{i}sica, Universidad Nacional Aut\'{o}noma de M\'{e}xico, Ciudad de M\'{e}xico, M\'{e}xico.
$^{b}$Facultad de Ingenier\'{i}a, Universidad Aut\'{o}noma de Quer\'{e}taro, Quer\'{e}taro, M\'{e}xico.

$^{c}$Centro de Investigaci\'{o}n en Computaci\'{o}n, Instituto Polit\'{e}cnico Nacional. Ciudad de M\'{e}xico, M\'{e}xico.

$^{d}$For a complete author list, see \color{blue}{www.hawc-observatory.org/collaboration/icrc2017.php}\\ 
}

\abstract{The HAWC observatory reconstructs atmospheric showers induced by very high-energy primary cosmic- or gamma-rays. Cosmic-rays with hadronic nature are several orders of magnitude more frequent than primary gammas and therefore it is necessary to perform the best gamma-hadron discrimination for each event to be able to survey the gamma-ray sky. Single vertical muons cause very characteristic signals in the 300 HAWC detectors. Particularly, those events that produce signals in the four PMTs of a tank corresponding to muons traversing the tank in the central region. In such cases exist the correlation that the largest signal is also the earliest to arrive, this due to the geometry of the Cherenkov light emission and the fact that relativistic muons travel faster than the light in the medium. This pattern is different from that of electrons, positrons and low energy gammas showering in the first meter of the tank water, which is typical of electromagnetic showers. In this work, we explored whether in air-showers it is possible to identify muon signals and whether this can help in the gamma/hadron discrimination of the recorded events.
          }

\FullConference{35th International Cosmic Ray Conference --- ICRC2017\\
		10--20 July, 2017\\
		Bexco, Busan, Korea}

\begin{document}

\section{Introduction}

\begin{figure}[H]
\includegraphics[scale=0.9]{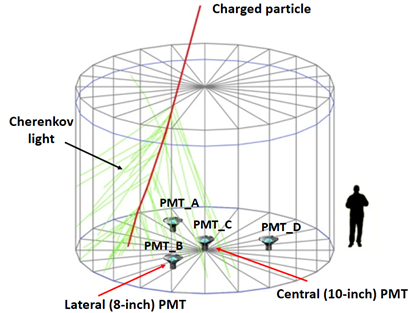} 
\centering
\caption{\label{Fig1} WCD setup, where the passage of a charged particle emitting Cherenkov photons is illustrated. }
\end{figure}

The High-Altitude Water Cherenkov (HAWC) observatory is an instrument located 4100 m a.s.l in Puebla, Mexico, with the mission of studying gamma-rays up to the very-high-energy domain (E > 100 GeV). The detector is an array of 300 Water Cherenkov Detectors (WCDs) or tanks, each one 4.5 m high, 7.3 m diameter and a volume containing 200 000 L of purified water. There are also four photomultiplier tubes (PMTs) anchored at the bottom of each tank: a high-quantum efficiency detector or PMT$\_$C (10-inch diameter) centrally located, surrounded by three equally spaced (8-inch diameter) PMTs (A, B, D); all upward facing (Fig.\,\ref{Fig1}) \cite{Smith}. 

The operating principle of HAWC is to sample air-shower particles by recording the Cherenkov light produced when they pass through the water. This coherent radiation is emitted into a forward cone (opening angle $\sim$ 41$^{\circ}$) that surrounds the direction of motion of the charged particle. Because the Cherenkov cone in water is so large, nearly every charged particle that enters the tank should be observed by at least one of the four PMTs, which converts the light into an electrical signal measurable by the data acquisition system. The output data corresponding to charge and time of the triggered PMTs, is later used for the shower's parameters reconstruction (e.g. core position, arrival direction and energy of the primary, event classification: gamma or hadron). 

A common issue of experiments like this is dealing with a large background originated by hadronic primary cosmic-rays. Such noise can be thousands of times greater than gamma signals (air-shower trigger rate in HAWC: $\sim$ 25 kHz), hence new methods for gamma/hadron separation are needed in order to increase the HAWC's sensitivity to new gamma-ray sources. 
Motivated by the fact that muon content is quite different in gamma-induced (poor in muons) and hadronic-induced (rich in muons) air-showers \cite{Grieder, Shapiro,Stanev}, we study the idea of formulating a new variable for background reduction related with counting the number of muons candidates present in the showers. To reach that goal, we first identified the signature of muons using only time information between photomultiplier tubes.

\section{Muon time signature in HAWC}

 Muons are the most abundant charged particle at the HAWC altitude and arrive mainly with some GeV of energy. They are originated from the charged meson decays ($\pi$, K) in air-showers, following approximately straight trajectories through the atmosphere and even inside WCD. Depending on the area where a muon travels it can trigger a specific number of PMTs, this is called multiplicity (M). In this work, we concentrate on vertical muons which travel close to the central light-sensor (PMT$\_$C), the region with the highest probability of having four multiplicities. 

\subsection{Analysis of single muons in raw data}

\begin{figure}[H]
\includegraphics[scale=0.27]{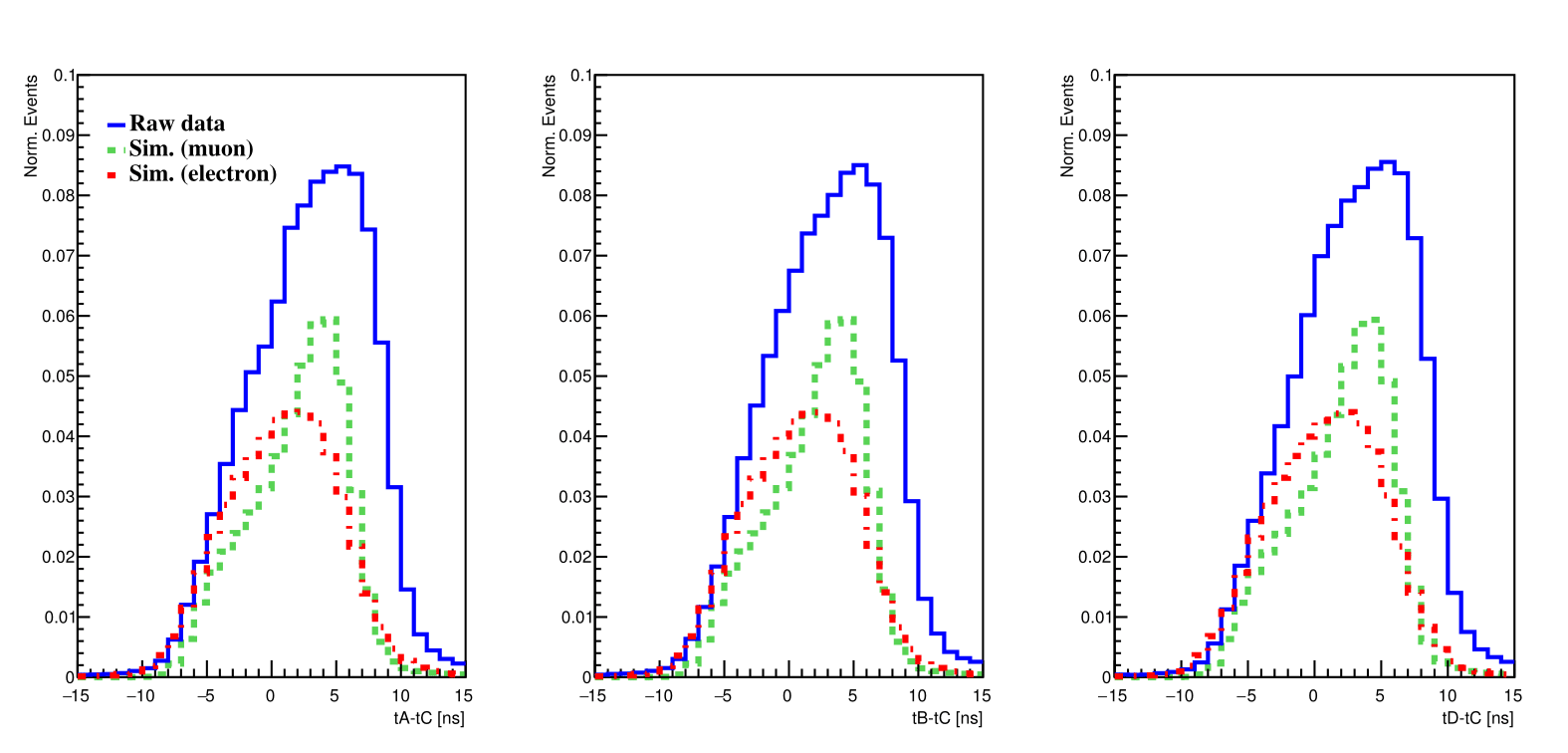} 
\centering
\caption{\label{Fig2} Comparison of the time difference distributions between PMTs for electrons (red), muons (green) and raw data (blue). }
\end{figure}

To choose muon candidates, first, we used a sample of raw data. This refers to the processed signal of the hit PMTs before the multiplicity trigger criterion for air-showers selection is applied. We considered all those tanks with four detections (M = 4) within a 15 ns time window. The pre\-vious selection might include the signal of electrons (e$^{-}$), positrons (e$^{+}$) and low energy gammas ($\gamma$) which triggered as well all 4-PMTs inside a WCD but to help in differentiation we perform Monte Carlo simulations of vertical electrons and muons in a single tank. The energy for particles was: electrons with 1 - 500 MeV and muons with 1 - 15 GeV. Then we analyzed the time difference distributions of the PMT$\_$C re\-lative to the peripherals one (t$_{A}$ - t$_{C}$, t$_{B}$ - t$_{C}$, t$_{D}$ - t$_{C}$). Figure\,\ref{Fig2} shows that the raw data and vertical muon distribution present a similar shape. The mean value (displaced to the right side = time C often smaller) is approximately 5 ns, a characteristic time in which the laterals PMTs see the Cherenkov light after the central detector does. However, the electron distribution is different with a time difference average value near 0 ns, meaning that the PMTs see the light practically at the same time. This happens because electrons stop quickly in the first meter of the water and then the light travels to the bottom falling promptly over all the detectors. On the contrary, muons reach the bottom triggering the PMT$\_$C first and later the lateral PMTs.

\subsection{Analysis in showers data}

Like gamma showers are mainly electromagnetic (containing a lot of e$^{-}$, e$^{+}$ and $\gamma$), they must exhibit a time difference close to 0 ns. In order to test this hypothesis, we compared simulation data of gamma-induced showers developed by the HAWC collaboration and real showers data which passed the trigger condition ( $>$ 28 hit PMTs within 150 ns). 

\begin{figure}[H]
\includegraphics[scale=0.75]{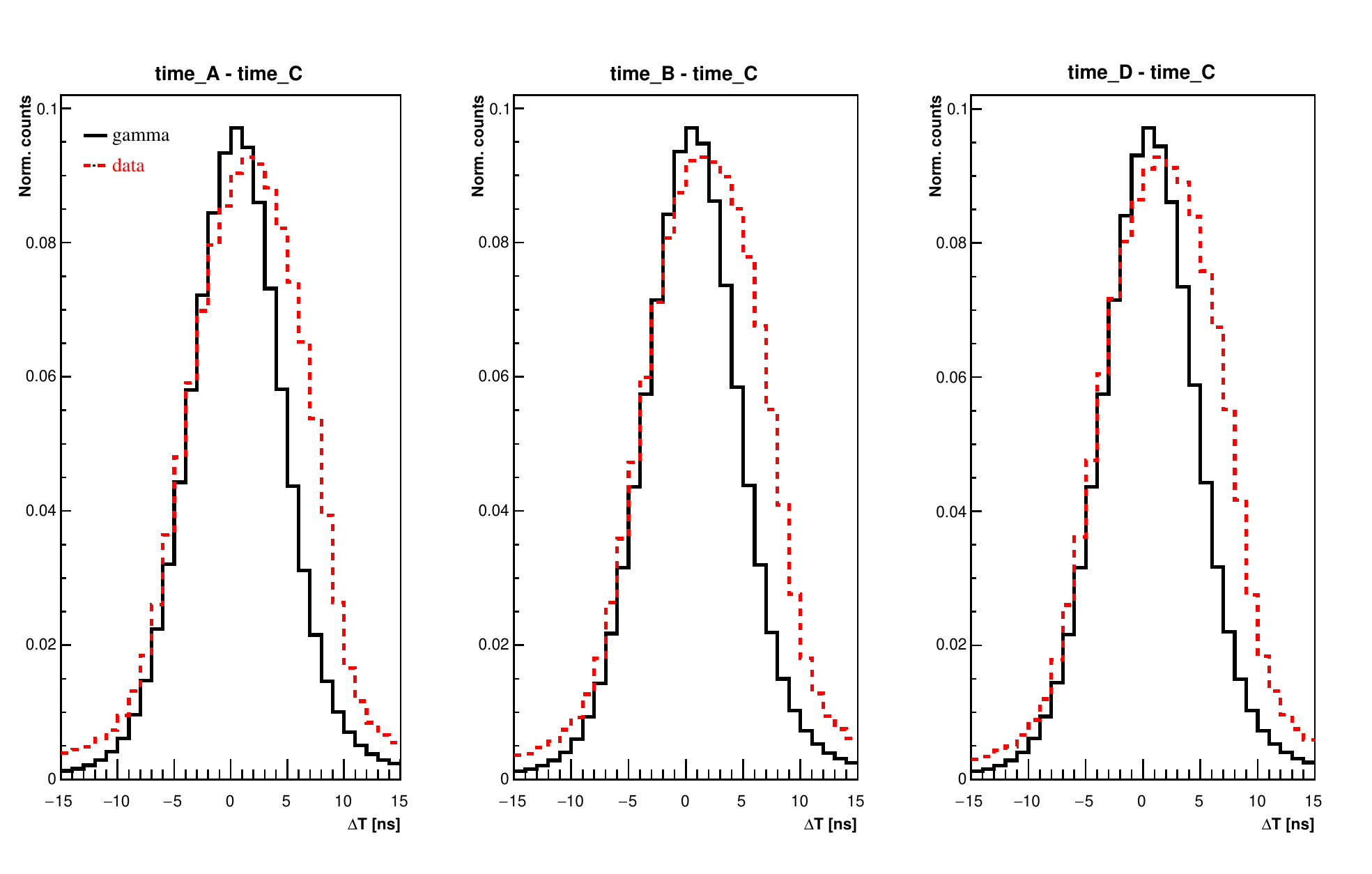} 
\centering
\caption{\label{Fig3} Comparison, using time difference distributions, of showers data (red) and gamma simulated data (black) for tanks with multiplicity 4.}
\end{figure}

In the simulation, the energy of gamma primaries range: 5 GeV - 500 TeV and their arrival direction: 0 - 75 degrees from zenith. Such data have an extended format, allowing to use the time of each triggered PMT in a shower. Likewise, the information of real cascades was extended data and they are presumably background due to their huge rate. 
We set the same restrictions to both kind of data (core inside the HAWC array, all available WCDs with M = 4, nearby vertical ($\pm$10$^{\circ}$) showers and time differences with respect to PMT$\_$C). 
Figure\,\ref{Fig3} shows the time difference distributions produced by showers with unequal muon content. The shower data is rich in muons, this is the reason why the time differences tend to about 5 ns.

\begin{figure}[H]
\includegraphics[scale=0.35]{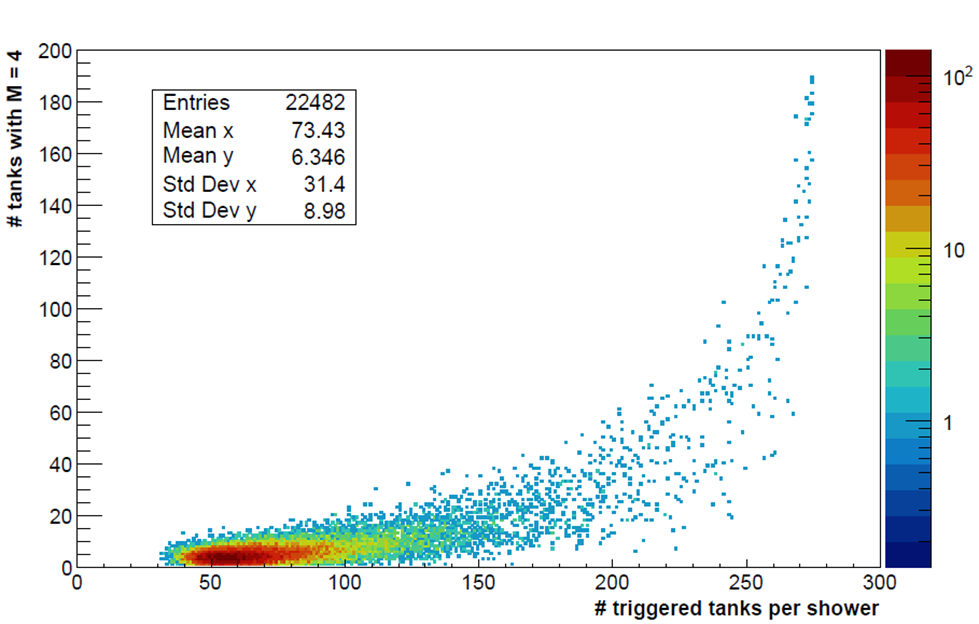} 
\centering
\caption{\label{Fig4} Number of tanks with multiplicity 4 as a function of the shower size (\# triggered tanks).}
\end{figure}

Since the selection of muon candidates is based on identifying the tanks with M = 4, it is extremely important to have a notion of the frequency they have per event. We determined this frequency using data (Fig.\,\ref{Fig4}). In the previous figure, most of the cascades always hit between 45 and 70 tanks. Furthermore, the number of WCDs with M = 4 increase with the size of the showers, but the periodicity of such events is very low. It is common for cascades of up to 150 tanks to have less than 20 tanks with M = 4.

\section{Definition of a new muon-content-based discrimination parameter}

\begin{figure}[H]
 \centering
  \includegraphics[width=0.35\textwidth]{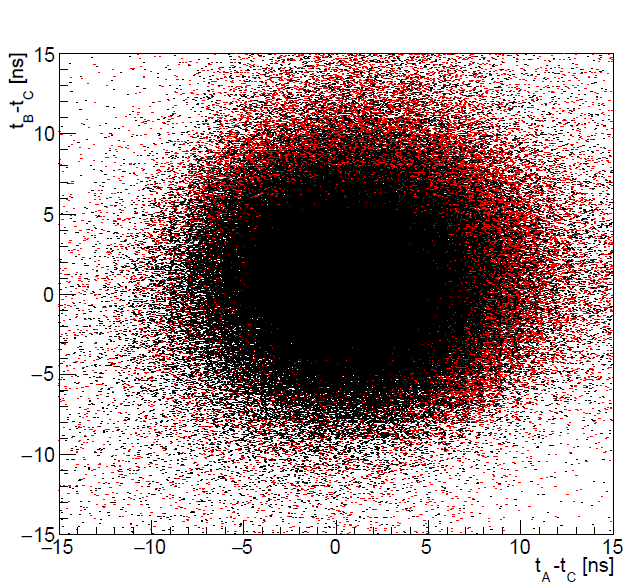}
  \includegraphics[width=0.35\textwidth]{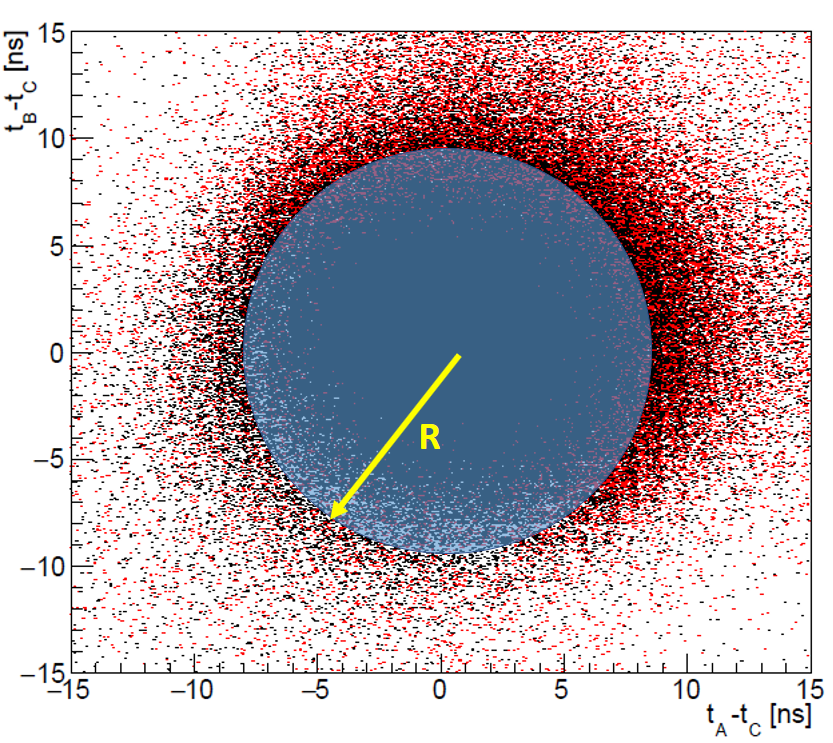}
 \caption{Left: two-dimensional plot created with two time differences, it includes the signal of simulated gamma (black dots) and real showers (red dots). Right: region of the cutting sphere with a defined radius \textbf{R}. }
 \label{Fig5}
\end{figure}

By making an overlap of the signals with two time differences (Fig.\,\ref{Fig5} - left), it can be seen that gammas (black dots) have a circular concentration whose center is close to zero, whereas the noise data (red dots) has an elliptical form, spread towards more large values in time difference. The idea of the variable born out of agglutinating the coincident data within a cutting sphere with a specific radius \textbf{R} (Fig.\,\ref{Fig5} - right). All those red dots outside the sphere are the signals occasioned by muons. Each WCD with M = 4 is a point in the space of the three time differences at a distance \textbf{$r_{i}$} from origin:  

\begin{equation}
 r_{i} = \sqrt{(t_{A} - t_{C})^{2} + (t_{B} - t_{C})^{2} + (t_{D} - t_{C})^{2} }
\end{equation}

\begin{figure}[H]
\includegraphics[scale=0.6]{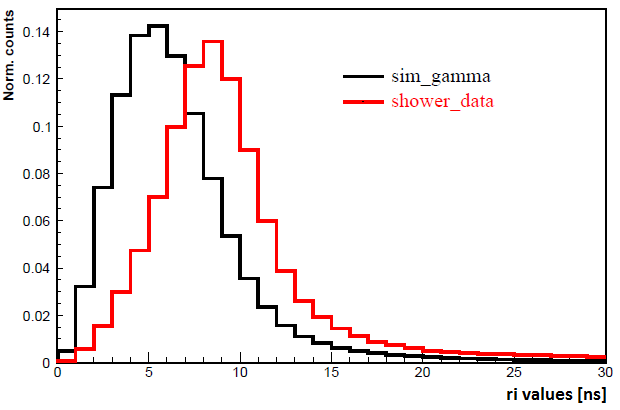} 
\centering
\caption{\label{Fig6} Histogram of \textbf{$r_{i}$} for both gamma and real-showers data.}
\end{figure}

In Fig.\,\ref{Fig6} it is shown the distribution of \textbf{$r_{i}$} for simulated gamma and measured hadronic showers. The value of \textbf{R} was set to 7 ns, such cut maximizes the number of muons out. Therefore, using the restriction \textbf{$r_{i}$} $>$ 7 ns were picked out the muon candidates. The new discrimination parameter named MUDECIDER, records the number of times the condition is met per event (overall number of muons) and its higher values correspond to hadronic-induced showers.  


\begin{figure}[H]
\includegraphics[scale=0.45]{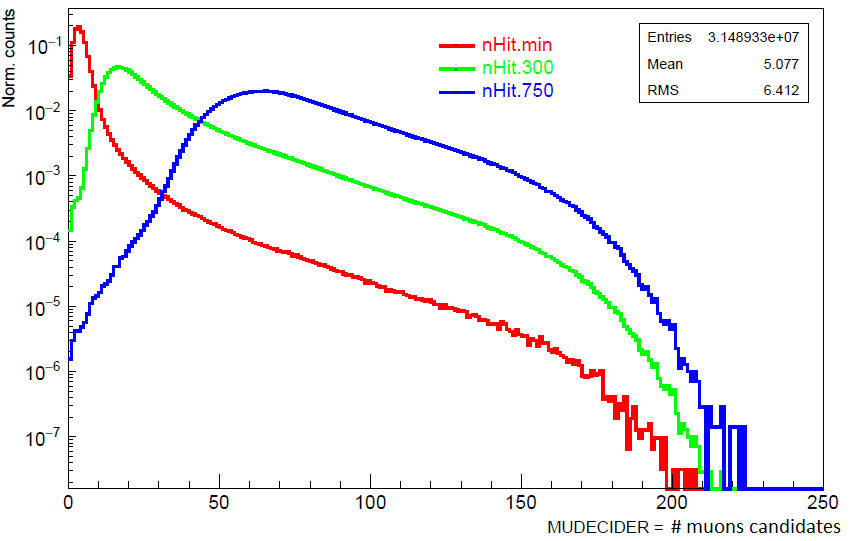} 
\centering
\caption{\label{Fig7} Distributions of MUDECIDER output values as a function of \textit{nHit}.}
\end{figure}

Adding a new project to the HAWC software the variable was implemented. Three types of reconstructed data which include the output information of MUDECIDER were generated. Cutting with the \textit{nHit} parameter (number of PMTs with signal within the trigger window), we obtained a normal reconstruction with the minimum threshold established by HAWC ($>$ 28 hit PMTs) and two others corresponding to showers where recorded Cherenkov light more than 300 and 750 PMTs (Fig.\,\ref{Fig7}). The bunches of data allowed to understand the variable's behavior as a function of larger and energetic events.

It should be noted that increasing \textit{nHit} more muon candidates are registered, the maximum number seen range between 200 - 230 candidates per shower. 

In order to verify MUDECIDER effectiveness, significance maps of the Crab Nebula were compared with and without using it. This astronomical object is the standard candle in TeV gamma-ray astronomy and HAWC observes it daily.    

\begin{figure}[H]
 \centering
  \includegraphics[width=0.47\textwidth]{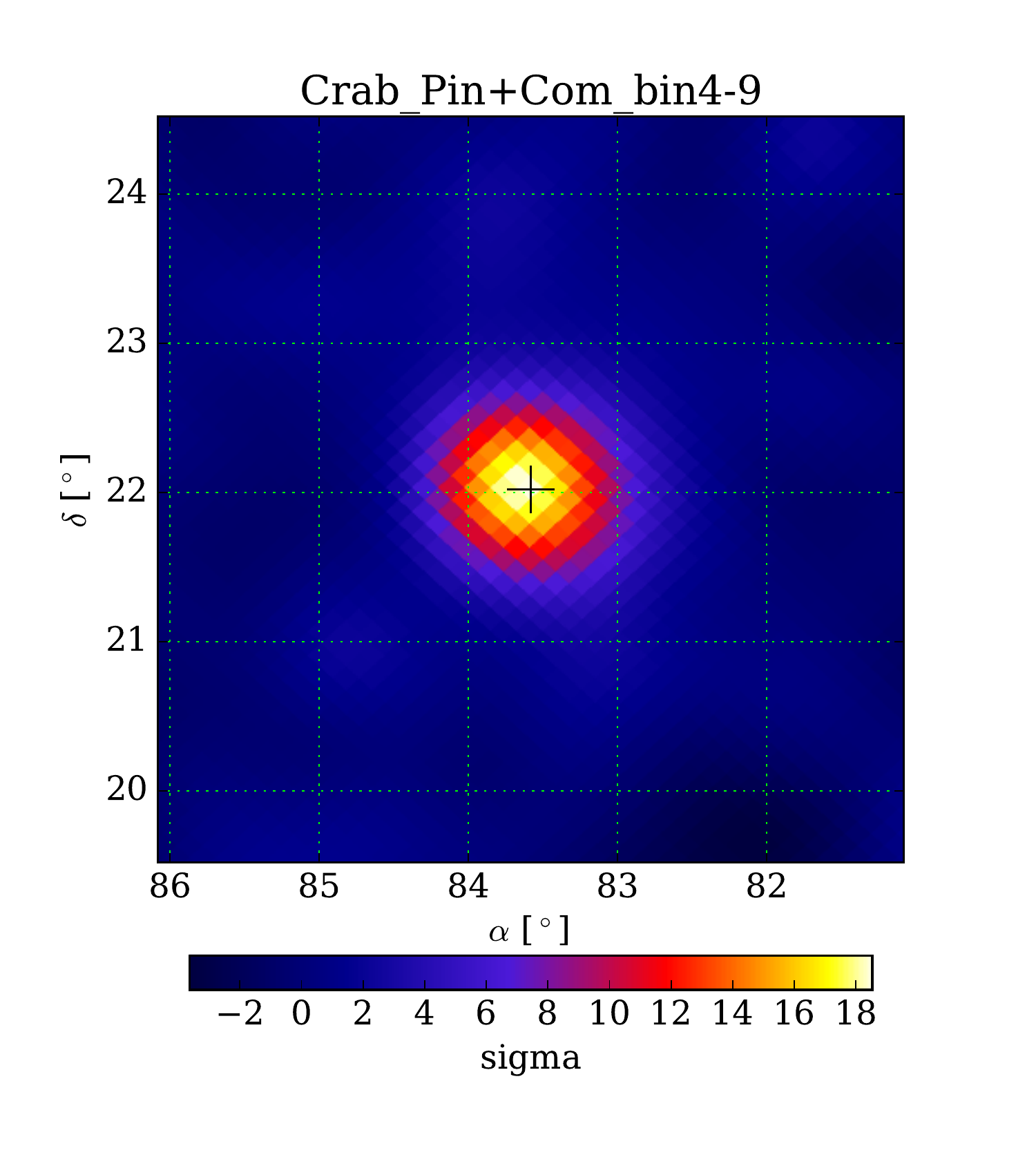}
  \includegraphics[width=0.47\textwidth]{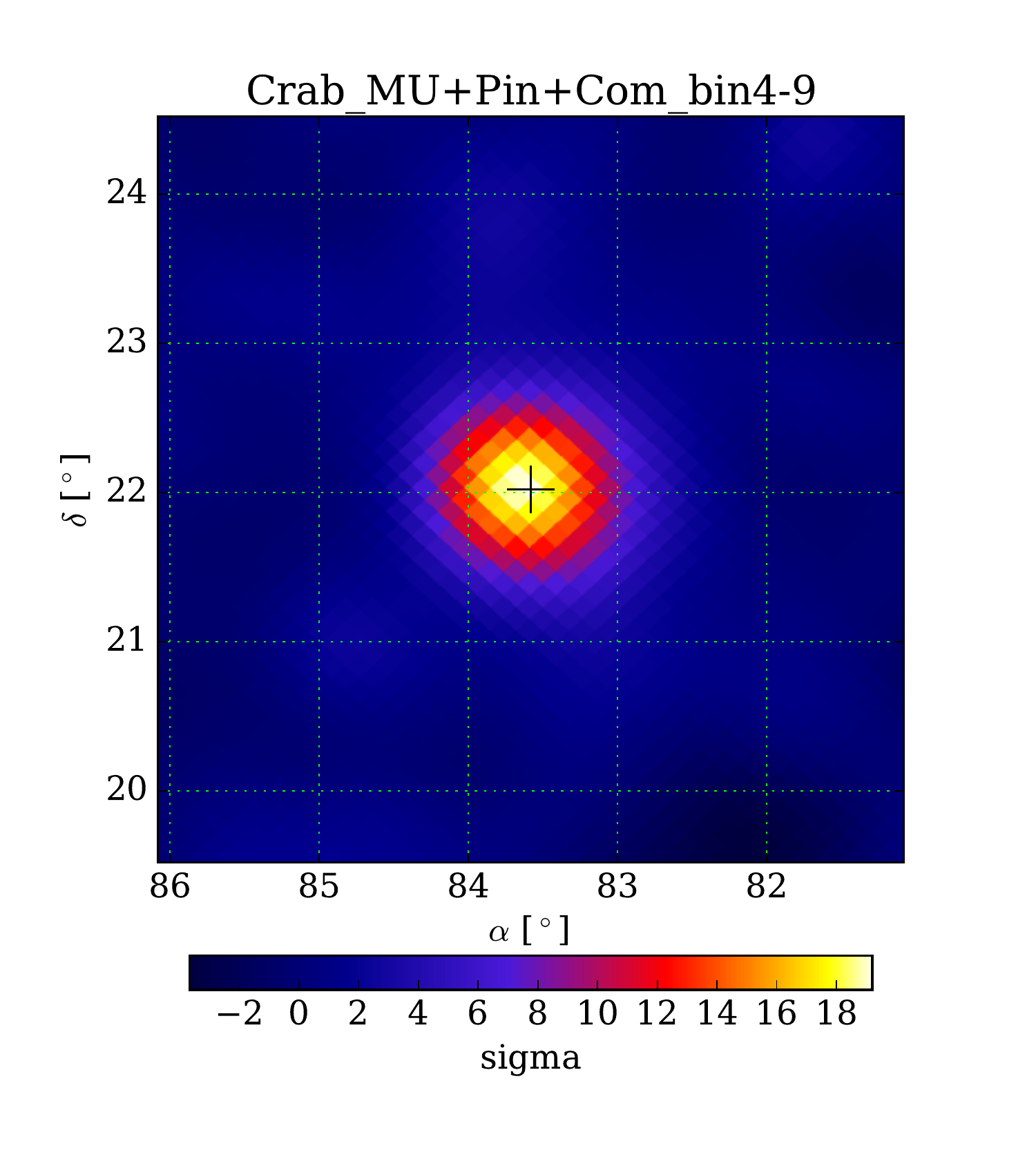}
 \caption{Significance maps of the Crab Nebula without (left) and using (right) MUDECIDER. There is a slight increase in the signal employing the new variable. }
 \label{Fig8}
\end{figure}

For the analysis in HAWC, cascade events are classified by their size into nine bins \cite{crab}, depending on the fraction of PMTs registering light in the detector.
To visualize the Crab, the second set of reconstructed data was employed (\textit{nHit} $>$ 300) and correspond to the bins 4 - 9. A first map (Fig.\,\ref{Fig8} - left) was created using the official gamma filters, PINCness (Pin) and Compactness (Com); later a second map including MUDECIDER (MU) was built (Fig.\,\ref{Fig8} - right).    

In the normal map (Pin + Com) the maximum significance was 18.53\,$\sigma$ and the minimum was -3.63\,$\sigma$; on the other hand, when the variable (MU + Pin + Com) was included, a maximum signi\-ficance of 19.20\,$\sigma$ was achieved and the minimum value was -3.66\,$\sigma$. For both maps, the maximum was found in the position (83.58$^{\circ}$, 22.02$^{\circ}$) and the minimum in (82.18$^{\circ}$, 19.63$^{\circ}$). The increase in significance was 0.67\,$\sigma$ and represents approximately 3.5\%. The maps have a duration of 555.28 hours and comprise the information of the bins together. It must be pointed out that the map with MUDECIDER contains fewer events. The cut values of the variable were established per bin analyzing the amount of signal/background in comparison to those produced in the bins of the normal map. The next step is to perform the optimization process of MUDECIDER. 

\section{Conclusion}

In the presented work, we analyzed the trail of vertical muons in the HAWC array using only time information. By performing time differences between photomultiplier tubes, we looked for muons with trajectories near the central area of tanks which induce multiplicity 4. Normally, they bring out a characteristic time difference of $\sim$ 5 ns between the PMT$\_$C relative to those periphe\-rals, this due to the well-known geometry of the Cherenkov light emission and the arrangement of the PMTs in the WCD. Then it was defined a new (under study) gamma/hadron separation variable based on the muon content of air-showers. Visualizing significance maps of the Crab Nebula, we showed promising results of this variable.

\section{Acknowledgments}

\justify We acknowledge the support from: the US National Science Foundation (NSF); the US Department of Energy Office of High-Energy Physics; the Laboratory Directed Research and Development (LDRD) program of Los Alamos National Laboratory; Consejo Nacional de Ciencia y Tecnolog\'{\i}a (CONACyT), M{\'e}xico (grants 271051, 232656, 260378, 179588, 239762, 254964, 271737, 258865, 243290, 132197), Laboratorio Nacional HAWC de rayos gamma; L'OREAL Fellowship for Women in Science 2014; Red HAWC, M{\'e}xico; DGAPA-UNAM (grants IG100317, IN111315, IN111716-3, IA102715, 109916, IA102917); VIEP-BUAP; PIFI 2012, 2013, PROFOCIE 2014, 2015; the University of Wisconsin Alumni Research Foundation; the Institute of Geophysics, Plan\-etary Physics, and Signatures at Los Alamos National Laboratory; Polish Science Centre grant DEC-2014/13/B/ST9/945; Coordinaci{\'o}n de la Investigaci{\'o}n Cient\'{\i}fica de la Universidad Michoacana. Thanks to Luciano D\'{\i}az and Eduardo Murrieta for technical support.


\begin{thebibliography}{99}

\bibitem{Smith}
\textbf{HAWC} Collaboration, A. Smith, HAWC: Design, Operation, Reconstruction and Analysis, in Proc. 34th ICRC, (The Hague, The Netherlands), August 2015.


\bibitem{Grieder}
P. Grieder. \textit{Extensive Air Showers}. In: Gamma Ray and Electron Initiated Air Showers (pp. 20). Springer-Verlag Berlin Heidelberg, 2010. eBook ISBN: 978-3-540-76941-5. 
 
\bibitem{Shapiro}
M.M. Shapiro, T. Stanev, and J.P. Wefel, editors. \textit{Astrophysical Sources of High Energy Particles
and Radiation}. Springer Science+Business Media Dordrecht, 2001. Volume 44 of Series II: Mathematics, Physics and Chemistry.(pp. 177 - 178)

\bibitem{Stanev}

T. Stanev.\textit{ Are deep underground detectors good $\gamma$-ray telescopes?}, Physical Review D. \textbf{33:9}, pp. 2740-2743, 1986.  

\bibitem{crab}
A. Abeysekara et al., \textit{Observation of the Crab Nebula with the HAWC Gamma-Ray Observatory}, \textit{The Astrophysical Journal}. \textbf{843:39} (17pp), 2017.
\end{thebibliography}
\end{document}